\input amssym.def
\input epsf


\magnification=\magstephalf
\hsize=14.0 true cm
\vsize=19 true cm
\hoffset=1.0 true cm
\voffset=2.0 true cm

\abovedisplayskip=12pt plus 3pt minus 3pt
\belowdisplayskip=12pt plus 3pt minus 3pt
\parindent=1.0em


\font\sixrm=cmr6
\font\eightrm=cmr8
\font\ninerm=cmr9

\font\sixi=cmmi6
\font\eighti=cmmi8
\font\ninei=cmmi9

\font\sixsy=cmsy6
\font\eightsy=cmsy8
\font\ninesy=cmsy9

\font\sixbf=cmbx6
\font\eightbf=cmbx8
\font\ninebf=cmbx9

\font\eightit=cmti8
\font\nineit=cmti9

\font\eightsl=cmsl8
\font\ninesl=cmsl9

\font\sixss=cmss8 at 8 true pt
\font\sevenss=cmss9 at 9 true pt
\font\eightss=cmss8
\font\niness=cmss9
\font\tenss=cmss10

 at 12 true pt
\font\bigrm=cmr10 at 12 true pt
 at 12 true pt

 at 14 true pt
\font\Bigrm=cmr12 at 16 true pt
 at 14 true pt

\catcode`@=11
\newfam\ssfam

\def\tenpoint{\def\rm{\fam0\tenrm}%
    \textfont0=\tenrm \scriptfont0=\sevenrm \scriptscriptfont0=\fiverm
    \textfont1=\teni  \scriptfont1=\seveni  \scriptscriptfont1=\fivei
    \textfont2=\tensy \scriptfont2=\sevensy \scriptscriptfont2=\fivesy
    \textfont3=\tenex \scriptfont3=\tenex   \scriptscriptfont3=\tenex
    \textfont\itfam=\tenit                  \def\it{\fam\itfam\tenit}%
    \textfont\slfam=\tensl                  \def\sl{\fam\slfam\tensl}%
    \textfont\bffam=\tenbf \scriptfont\bffam=\sevenbf
    \scriptscriptfont\bffam=\fivebf
                                            \def\bf{\fam\bffam\tenbf}%
    \textfont\ssfam=\tenss \scriptfont\ssfam=\sevenss
    \scriptscriptfont\ssfam=\sevenss
                                            \def\ss{\fam\ssfam\tenss}%
    \normalbaselineskip=13pt
    \setbox\strutbox=\hbox{\vrule height8.5pt depth3.5pt width0pt}%
    \let\big=\tenbig
    \normalbaselines\rm}

\def\ninepoint{\def\rm{\fam0\ninerm}%
    \textfont0=\ninerm      \scriptfont0=\sixrm
                            \scriptscriptfont0=\fiverm
    \textfont1=\ninei       \scriptfont1=\sixi
                            \scriptscriptfont1=\fivei
    \textfont2=\ninesy      \scriptfont2=\sixsy
                            \scriptscriptfont2=\fivesy
    \textfont3=\tenex       \scriptfont3=\tenex
                            \scriptscriptfont3=\tenex
    \textfont\itfam=\nineit \def\it{\fam\itfam\nineit}%
    \textfont\slfam=\ninesl \def\sl{\fam\slfam\ninesl}%
    \textfont\bffam=\ninebf \scriptfont\bffam=\sixbf
                            \scriptscriptfont\bffam=\fivebf
                            \def\bf{\fam\bffam\ninebf}%
    \textfont\ssfam=\niness \scriptfont\ssfam=\sixss
                            \scriptscriptfont\ssfam=\sixss
                            \def\ss{\fam\ssfam\niness}%
    \normalbaselineskip=12pt
    \setbox\strutbox=\hbox{\vrule height8.0pt depth3.0pt width0pt}%
    \let\big=\ninebig
    \normalbaselines\rm}

\def\eightpoint{\def\rm{\fam0\eightrm}%
    \textfont0=\eightrm      \scriptfont0=\sixrm
                             \scriptscriptfont0=\fiverm
    \textfont1=\eighti       \scriptfont1=\sixi
                             \scriptscriptfont1=\fivei
    \textfont2=\eightsy      \scriptfont2=\sixsy
                             \scriptscriptfont2=\fivesy
    \textfont3=\tenex        \scriptfont3=\tenex
                             \scriptscriptfont3=\tenex
    \textfont\itfam=\eightit \def\it{\fam\itfam\eightit}%
    \textfont\slfam=\eightsl \def\sl{\fam\slfam\eightsl}%
    \textfont\bffam=\eightbf \scriptfont\bffam=\sixbf
                             \scriptscriptfont\bffam=\fivebf
                             \def\bf{\fam\bffam\eightbf}%
    \textfont\ssfam=\eightss \scriptfont\ssfam=\sixss
                             \scriptscriptfont\ssfam=\sixss
                             \def\ss{\fam\ssfam\eightss}%
    \normalbaselineskip=10pt
    \setbox\strutbox=\hbox{\vrule height7.0pt depth2.0pt width0pt}%
    \let\big=\eightbig
    \normalbaselines\rm}

\def\tenbig#1{{\hbox{$\left#1\vbox to8.5pt{}\right.\n@space$}}}
\def\ninebig#1{{\hbox{$\textfont0=\tenrm\textfont2=\tensy
                       \left#1\vbox to7.25pt{}\right.\n@space$}}}
\def\eightbig#1{{\hbox{$\textfont0=\ninerm\textfont2=\ninesy
                       \left#1\vbox to6.5pt{}\right.\n@space$}}}

\font\sectionfont=cmbx10
\font\subsectionfont=cmti10

\def\figurecaptionfont{\ninepoint}
\def\tablecaptionfont{\ninepoint}
\def\footnotefont{\eightpoint}


\newcount\equationno
\newcount\bibitemno
\newcount\figureno
\newcount\tableno

\equationno=0
\bibitemno=0
\figureno=0
\tableno=0


\footline={\ifnum\pageno=0{\hfil}\else
{\hss\rm\the\pageno\hss}\fi}


\def\section #1. #2 \par
{\vskip0pt plus .20\vsize\penalty-100 \vskip0pt plus-.20\vsize
\vskip 1.6 true cm plus 0.2 true cm minus 0.2 true cm
\global\def\equationlabel{#1}
\global\equationno=0
\leftline{\sectionfont #1. #2}\par
\immediate\write\terminal{Section #1. #2}
\vskip 0.7 true cm plus 0.1 true cm minus 0.1 true cm
\noindent}


\def\subsection #1 \par
{\vskip0pt plus 0.8 true cm\penalty-50 \vskip0pt plus-0.8 true cm
\vskip2.5ex plus 0.1ex minus 0.1ex
\leftline{\subsectionfont #1}\par
\immediate\write\terminal{Subsection #1}
\vskip1.0ex plus 0.1ex minus 0.1ex
\noindent}


\def\appendix #1 \par
{\vskip0pt plus .20\vsize\penalty-100 \vskip0pt plus-.20\vsize
\vskip 1.6 true cm plus 0.2 true cm minus 0.2 true cm
\global\def\equationlabel{\hbox{\rm#1}}
\global\equationno=0
\leftline{\sectionfont Appendix #1}\par
\immediate\write\terminal{Appendix #1}
\vskip 0.7 true cm plus 0.1 true cm minus 0.1 true cm
\noindent}


\def\equation#1{$$\displaylines{\qquad #1}$$}
\def\enum{\global\advance\equationno by 1
\hfill\llap{(\equationlabel.\the\equationno)}}
\def\noenum{\hfill}
\def\next#1{\cr\noalign{\vskip#1}\qquad}


\def\ifundefined#1{\expandafter\ifx\csname#1\endcsname\relax}

\def\ref#1{\ifundefined{#1}?\immediate\write\terminal{unknown reference
on page \the\pageno}\else\csname#1\endcsname\fi}

\newwrite\terminal
\newwrite\bibitemlist

\def\bibitem#1#2\par{\global\advance\bibitemno by 1
\immediate\write\bibitemlist{\string\def
\expandafter\string\csname#1\endcsname
{\the\bibitemno}}
\item{[\the\bibitemno]}#2\par}

\def\beginbibliography{
\vskip0pt plus .15\vsize\penalty-100 \vskip0pt plus-.15\vsize
\vskip 1.2 true cm plus 0.2 true cm minus 0.2 true cm
\leftline{\sectionfont References}\par
\immediate\write\terminal{References}
\immediate\openout\bibitemlist=biblist
\frenchspacing\parindent=1.8em
\vskip 0.5 true cm plus 0.1 true cm minus 0.1 true cm}

\def\endbibliography{
\immediate\closeout\bibitemlist
\nonfrenchspacing\parindent=1.0em}


\def\figurecaption#1{\global\advance\figureno by 1
\narrower\figurecaptionfont
Fig.~\the\figureno. #1}

\def\tablecaption#1{\global\advance\tableno by 1
\vbox to 0.5 true cm { }
\centerline{\tablecaptionfont%
Table~\the\tableno. #1}
\vskip-0.4 true cm}

\def\thintablerule{\hrule height0.4pt}

\tenpoint

\immediate\openin\bibitemlist=biblist
\ifeof\bibitemlist\immediate\closein\bibitemlist
\else\immediate\closein\bibitemlist
\input biblist \fi


\def\thismonth{\ifcase\month\or
January\or February\or March\or April\or May\or June\or
July\or August\or September\or October\or November\or December\fi}



\def\rmd{{\rm d}}
\def\rmD{{\rm D}}
\def\rme{{\rm e}}
\def\rmO{{\rm O}}
\def\rmU{{\rm U}}


\def\gz{{\Bbb Z}}

\def\Re{{\rm Re}\,}


\def\proof{\noindent{\sl Proof:}\kern0.6em}

\def\frac#1#2{\hbox{$#1\over#2$}}
\def\dual{\mathstrut^*\kern-0.1em}

\def\ring{\mathaccent"7017}
\def\lvec#1{\setbox0=\hbox{$#1$}
    \setbox1=\hbox{$\scriptstyle\leftarrow$}
    #1\kern-\wd0\smash{
    \raise\ht0\hbox{$\raise1pt\hbox{$\scriptstyle\leftarrow$}$}}
    \kern-\wd1\kern\wd0}
\def\rvec#1{\setbox0=\hbox{$#1$}
    \setbox1=\hbox{$\scriptstyle\rightarrow$}
    #1\kern-\wd0\smash{
    \raise\ht0\hbox{$\raise1pt\hbox{$\scriptstyle\rightarrow$}$}}
    \kern-\wd1\kern\wd0}


\def\nab#1{{\nabla_{#1}}}
\def\nabstar#1{{\nabla\kern0.5pt\smash{\raise 4.5pt\hbox{$\ast$}}
               \kern-5.5pt_{#1}}}
\def\drv#1{{\partial_{#1}}}
\def\drvstar#1{{\partial\kern0.5pt\smash{\raise 4.5pt\hbox{$\ast$}}
               \kern-6.0pt_{#1}}}

\def\ldrvstar#1{{\lvec{\,\partial}\kern-0.5pt\smash{\raise 4.5pt\hbox{$\ast$}}
               \kern-5.0pt_{#1}}}




\def\psibar{\overline{\psi}}


\def\dirac#1{\gamma_{#1}}
\def\diracstar#1#2{
    \setbox0=\hbox{$\gamma$}\setbox1=\hbox{$\gamma_{#1}$}
    \gamma_{#1}\kern-\wd1\kern\wd0
    \smash{\raise4.5pt\hbox{$\scriptstyle#2$}}}
\def\dirachat{\hat{\gamma}_5}


\def\group{G}
\def\algebra{{\frak g}}
\def\SUtwo{{\rm SU(2)}}

\def\tr{{\rm tr}}
\def\Tr{{\rm Tr}}
\def\Ad{{\rm Ad}\kern0.1em}
\def\d#1{d^{#1}_{\hbox{$\scriptstyle\kern-0.5pt R\scriptfont1=\sixi$}}}


\def\subF{{\vrule height6.0pt depth1.5pt width0pt}_{\rm F}}
\def\Seff{S_{\rm eff}}
\def\deltabar{\bar{\delta}}
\def\L{{\frak L}}
\def\F{{\frak F}}
\def\H{{\frak H}}
\def\Lcheck{{\check{\frak L}}}
\def\Fcheck{{\check{\frak F}}}
\def\Ahat{{\widehat{A}}}
\rightline{CERN-TH/2000-165}

\vskip 2.0 true cm 
\centerline
{\Bigrm  Lattice regularization of chiral gauge theories}
\vskip 1.6ex
\centerline
{\Bigrm  to all orders of perturbation theory}
\vskip 0.6 true cm
\centerline{\bigrm 
Martin L\"uscher\kern1pt%
\footnote{${\vrule height7.0pt depth1.5pt width0pt}^{\ast}$}{\footnotefont%
On leave from Deutsches Elektronen-Synchrotron DESY, 
D-22603 Hamburg, Germany}
}
\vskip3ex
\centerline{\it CERN, Theory Division} 
\centerline{\it CH-1211 Geneva 23, Switzerland}
\vskip 0.8 true cm
\thintablerule
\vskip 2.0ex
\ninepoint
\leftline{\bf Abstract}
\vskip 1.0ex\noindent
In the framework of perturbation theory, it is possible to put chiral
gauge theories on the lattice without violating the gauge symmetry or
other fundamental principles, provided the fermion representation of
the gauge group is anomaly-free.
The basic elements of this construction (which starts from the
Ginsparg--Wilson relation) are
briefly recalled and the exact
cancellation of the gauge anomaly, at any
fixed value of the lattice spacing and for any compact gauge group,
is then proved rigorously through a recursive procedure.

\vskip 2.0ex
\thintablerule
\vskip -3.0ex
\tenpoint

\section 1. Introduction

In chiral gauge theories the perturbation expansion is not easy to set
up consistently, because the widely used regularization methods (and
also the BPHZ finite-part prescription) violate the gauge symmetry.
Non-invariant counterterms must then be included in the action, with
coefficients chosen so as to restore the symmetry after
renormalization and removal of the regularization
[\ref{BRS}--\ref{PiguetSorella}].
As a consequence~the proof of the renormalizability of these theories
is far more complicated than in the case of ordinary gauge theories.
The complexity of the subtraction procedure also presents a difficulty
in practice when calculating higher-order corrections to electro\-weak
processes (see refs.~[\ref{GrassiEtAl},\ref{Jegerlehner}] for a recent
discussion and further references).

Essentially the same (regularization plus subtraction) strategy
can be adopted in 
lattice gauge theory, where it is referred to as 
the ``Rome approach" [\ref{Rome}--\ref{RomeReviewII}].
In this case the BRS invariance is 
broken by the Wilson term,
which is needed in the fermion action to 
avoid the infamous species-doubling problem.
The symmetry is then recovered in the continuum limit
after adding the appropriate counterterms.

In the present paper a different approach is described,
in which the regularization preserves the gauge invariance 
of the theory to all orders of the gauge coupling.
For many years this seemed to be excluded,
but after the rediscovery of the Ginsparg--Wilson relation
[\ref{GinspargWilson}--\ref{Niedermayer}]
the situation changed and a general formulation of 
chiral gauge theories on the lattice has emerged, where 
the cancellation of the symmetry-breaking 
terms {\it at any fixed value of the lattice spacing}\/ 
reduces to a local cohomology problem
[\ref{AbelianChLGT}--\ref{ReviewChLGT}].
The latter appears at the one-loop level, and once the 
symmetry is restored to this order of the loop expansion, 
the theory is guaranteed to be gauge-invariant 
at all higher orders too.

The cohomology problem was first solved for 
abelian gauge groups [\ref{AbelianCohomology},\ref{FujiwaraEtAl}],
and the general solution, to all orders of the gauge coupling
and for any compact gauge group, has recently
been obtained by Suzuki [\ref{SuzukiAnomaly}]%
\kern1.5pt%
\footnote{$\dagger$}{\footnotefont%
Beyond perturbation theory the solution of the
cohomology problem is known
for abelian gauge groups [\ref{AbelianCohomology},\ref{FujiwaraEtAl}]
and for $\SUtwo\times\rmU(1)$ [\ref{KikukawaNakayama}].
The consistent formulation of chiral lattice gauge theories
at this level also requires a proof of the absence of global topological
obstructions. So far this has only been achieved 
in the abelian case [\ref{AbelianChLGT}].
}.
In his paper Suzuki starts from the Wess--Zumino 
consistency condition and roughly follows the established
strategies in the continuum theory [\ref{StoraI}--\ref{Bertlmann}].
This turns out to be rather complicated,
but as will be shown here the exact anomaly cancellation
can also be proved in a more direct and significantly simpler way.

In the next section the formulation of chiral lattice gauge theories
along the lines of 
refs.~[\ref{AbelianChLGT}--\ref{ReviewChLGT}] 
is briefly recalled.
The Feynman rules in these theories (sect.~3) are essentially
as in lattice QCD, except for the chiral projectors in the 
fermion pro\-pa\-ga\-tor and
a set of additional local gauge field vertices
with five or more legs that constitute the solution of the cohomology
problem alluded to above.
In sect.~4 an algebraic proof of 
the existence of this solution is given, using 
the classification theorem for topological fields 
in abelian lattice gauge theories 
of refs.~[\ref{AbelianCohomology},\ref{FujiwaraEtAl}].
The paper ends with a short discussion of further results
and a few concluding remarks.

\section 2. Chiral lattice gauge theories

The chiral gauge theories discussed in this and the following
two sections
involve a multiplet of left-handed fermions 
but no Higgs fields or other matter fields, since
these can be easily included later if so desired.
For simplicity any details of the lattice formulation 
that are only relevant at the non-perturbative level
(global anomalies, for example
[\ref{Witten}--\ref{BaerCampos}]) 
will be skipped over without further notice.

\subsection 2.1 Fields and lattice action

As usual the theory is set up on a four-dimensional
euclidean lattice with spacing $a$.
The gauge group $\group$ is assumed to be a compact connected Lie
group, and the gauge field is represented by link variables
$U(x,\mu)\in\group$, where $x$ runs over all lattice points and
$\mu=0,\ldots,3$ labels the lattice axes.
We first consider lattice Dirac fields $\psi(x)$ that 
transform according to some unitary representation $R$ of the
gauge group and defer the discussion of how to eliminate the
right-handed components to the next subsection.

A key element of the present approach to chiral lattice gauge theories
is the choice of a lattice Dirac operator $D$
that satisfies the Ginsparg--Wilson relation [\ref{GinspargWilson}]
\equation{
  \dirac{5}D+D\dirac{5}=aD\dirac{5}D
  \enum
}
and the hermiticity condition $D^{\dagger}=\dirac{5}D\dirac{5}$.
The operator should also be local, gauge-covariant
and have a number of further properties [\ref{Locality},\ref{AbelianChLGT}], 
as any other acceptable lattice Dirac operator.
A relatively simple expression, which fulfils all
these require\-ments, is given by 
[\ref{NeubergerOperator}]
\equation{
  D={1\over a}\kern0.5pt\bigl\{1-A(A^{\dagger}A)^{-1/2}\bigr\},
  \qquad
  A=1-aD_{\rm w},
  \enum
}
where $D_{\rm w}$ denotes the standard Wilson--Dirac operator
\equation{
  D_{\rm w}=\frac{1}{2}\kern0.5pt\bigl\{\dirac{\mu}(\nabstar{\mu}+\nab{\mu})
  -a\nabstar{\mu}\nab{\mu}\bigr\}
  \enum
}
(see appendix A for unexplained notations).
In the following we shall stick to this operator, but it should
be emphasized that other acceptable
solutions of the Gins\-parg--Wilson relation would do just as well.
For perturbation theory it would actually be sufficient to 
provide a solution in the form of a formal power
series expansion in the gauge potential.

In the present context the standard plaquette action
\equation{
  S_{\rm G}[U]=
  {1\over g_0^2}\sum_x\sum_{\mu,\nu}\Re\tr\{1-P(x,\mu,\nu)\},
  \enum
  \next{2ex}
  P(x,\mu,\nu)=U(x,\mu)U(x+a\hat{\mu},\nu)
  U(x+a\hat{\nu},\mu)^{-1}U(x,\nu)^{-1},
  \enum
}
is a possible choice for the gauge field action,
with $g_0$ the bare coupling, and the fermion action
is taken to be of the usual form
\equation{
  S_{\rm F}[U,\psibar,\psi]=a^4\sum_x\psibar(x)D\psi(x).
  \enum
}
At this stage the theory thus looks like lattice QCD, 
except for the fact that we 
have allowed the fermions to be in an arbitrary representation $R$
of the gauge group.

\subsection 2.2 Chiral projection

An important consequence of the Ginsparg--Wilson relation is
that the fermion action admits 
an exact chiral symmetry [\ref{ExactSymmetry}]
that can be used to separate the chiral com\-ponents
of the fermion field in a natural way 
[\ref{Narayanan}--\ref{AbelianChLGT}].
One first observes that the operator
$\dirachat=\dirac{5}(1-aD)$ satisfies
\equation{
  (\dirachat)^{\dagger}=\dirachat,
  \qquad
  (\dirachat)^2=1, 
  \qquad
  D\dirachat=-\dirac{5}D.
  \enum
}
The fermion action 
thus splits into left- and right-handed parts if the chiral 
projectors for fermion and antifermion fields 
are defined through
\equation{ 
  \hat{P}_{\pm}=\frac{1}{2}(1\pm\dirachat),
  \qquad
  P_{\pm}=\frac{1}{2}(1\pm\dirac{5}),
  \enum
}
respectively. In particular, by imposing the constraints
\equation{
  \hat{P}_{-}\psi=\psi,
  \qquad
  \psibar P_{+}=\psibar,
  \enum
}
the right-handed components are eliminated
and one obtains a chiral gauge theory that
is completely consistent at the classical level.

\subsection 2.3 Correlation functions

Expectation values of products ${\cal O}$ of the basic fields
are defined through the functional integral
\equation{
  \langle{\cal O}\rangle={1\over{\cal Z}}
  \int\rmD[U]
  \rmD[\kern1pt\psi\kern1pt]\rmD[\kern1pt\psibar\kern1pt]\,
  {\cal O}\kern1pt
  \rme^{-S_{\rm G}[U]-S_{\rm F}[U,\psibar,\psi]},
  \enum
}
where $\rmD[U]$ denotes the standard integration measure for 
lattice gauge fields and the normalization factor ${\cal Z}$ is 
defined through the requirement that $\langle1\rangle=1$.
The fermion integral should be restricted to the subspace of left-handed
fields, which can be easily done for any given
gauge field configuration.
However, since the subspace of left-handed fields changes 
with the gauge field, there is no obvious way 
to fix the relative phase of the fermion integration measure
at different points in field space.
As a consequence the fermion partition function
\equation{
  \rme^{-\Seff[U]}=
  \int
  \rmD[\kern1pt\psi\kern1pt]\rmD[\kern1pt\psibar\kern1pt]\,
  \rme^{-S_{\rm F}[U,\psibar,\psi]}
  \enum
}
has a non-trivial
phase ambiguity [\ref{AbelianChLGT}--\ref{ReviewChLGT}].

Apart from this, the fermion integral is well-defined and of the 
Gaussian type. Equation (2.10) can thus be rewritten in the form
\equation{
  \langle{\cal O}\rangle=
  {1\over{\cal Z}}
  \int \rmD[U]\,\{{\cal O}\}\subF
  \kern1pt\rme^{-S_{\rm G}[U]-S_{\rm eff}[U]},
  \enum
}  
where $\{{\cal O}\}\subF$ is a sum of Wick contractions
that are obtained by applying Wick's theorem to the 
fermion fields in $\cal O$ and substituting
\equation{
  \{\psi(x)\psibar(y)\}\subF=
  \hat{P}_{-}S(x,y)P_{+},
  \qquad
  DS(x,y)=a^{-4}\delta_{xy},
  \enum
}
for the basic contraction. Note the presence of the 
chiral projectors in this formula, which make it explicit that the 
propagating fermions are left-handed.

\subsection 2.4 Measure term and gauge invariance

To complete the definition of the lattice theory,
we now need to say how
the phase of the fermion measure is to be determined.
Since only the relative phase at different points in field
space matters, the problem may be approached
by computing the change of the effective action
$\Seff$ under variations 
\equation{
  \delta_{\eta}U(x,\mu)=a\eta_{\mu}(x)U(x,\mu),
  \qquad
  \eta_{\mu}(x)=\eta^a_{\mu}(x)T^a,
  \enum
}
of the link field (cf.~appendix A).
Apart from the naively expected term,
the result of this calculation
[\ref{AbelianChLGT}--\ref{ReviewChLGT}]
\equation{
  \delta_{\eta}\Seff=-\Tr\{\delta_{\eta}D\hat{P}_{-}D^{-1}P_{+}\}+
  i\L_{\eta}
  \enum
}
involves a second term (the {\it measure term})
that arises from the implicit dependence of 
the fermion measure on the gauge field.
$\L_{\eta}$ is linear in the field variation,
\equation{
  \L_{\eta}=a^4\sum_x\eta^a_{\mu}(x)j^a_{\mu}(x),
  \enum
}
where $j_{\mu}(x)$ is a function of the gauge field
that contains all the non-trivial information about 
the phase of the fermion measure.

At this point little is known about this current,
but it is straightforward to write down a few general requirements 
that turn out to be very restrictive and 
essentially fix the phase of the measure
[\ref{AbelianChLGT}--\ref{ReviewChLGT}].
In perturbation theory 
the situation is particularly simple, and
the only conditions that must be fulfilled are%
\kern1.5pt\footnote{$\dagger$}{\footnotefont%
The notion of locality which is being used here 
is the same as in the earlier work on the subject 
(see ref.~[\ref{AbelianCohomology}], for example).
What precisely this means in perturbation theory is explained
in sect.~3.}

\vskip1ex
\noindent
(a) {\sl The current $j_{\mu}(x)$ is a gauge-covariant local field.} 

\vskip1ex
\noindent
(b) {\sl $\L_{\eta}$ satisfies the 
integrability condition
\equation{
  \delta_{\eta}\L_{\zeta}-\delta_{\zeta}\L_{\eta}+
  a\L_{[\eta,\zeta]}=
  i\kern1pt
  \Tr\{\hat{P}_{-}[\delta_{\eta}\hat{P}_{-},\delta_{\zeta}\hat{P}_{-}]\}
  \rm\enum
}
for all field variations $\eta_{\mu}(x)$ and
$\zeta_{\mu}(x)$ that do not depend on the gauge field.}

\vskip1ex
\noindent
The measure term $\L_{\eta}$ may then be interpreted 
as a local counterterm, 
which has to be included to ensure the integrability  
of the right-hand side of eq.~(2.15).
Since the existence of an underlying fermion measure 
is guaranteed if (b) holds [\ref{NonAbelianChLGT}], 
one can in fact
{\it define the theory}\/ through eqs.~(2.12)--(2.16),
with some particular choice of the current $j_{\mu}(x)$ that
satisfies conditions (a) and (b).

In the following we adopt this point of view
and shall show (in sect.~4) that 
such a current can be constructed to all orders
of the gauge coupling
if the fermion multiplet is anomaly-free, 
i.e.~if the tensor 
\equation{
  \d{abc}=2i
  \kern1pt\tr\bigr\{R(T^a)[R(T^b)R(T^c)+R(T^c)R(T^b)]\bigl\}
  \enum
}
vanishes. There are no further restrictions on the fermion representation
or the gauge group, and the solution is
unique up to irrelevant local terms that amount 
to a redefinition of the lattice action of the gauge field.

Once the phase ambiguity has been fixed in this way, 
the effective action $\Seff$ (and thus the whole theory) 
can be shown to be gauge-invariant.
The proof of this important result is given in 
appendix B. 
Here we only note that
infinitesimal gauge transformations are generated by lattice fields
$\omega(x)$ with values in the Lie algebra of~$\group$. The corresponding
variations of the link variables are obtained by substituting
\equation{
  \eta_{\mu}(x)=-\nab{\mu}\omega(x)
  \enum
}
in eq.~(2.14), and the gauge invariance of the effective action is 
then equivalent to 
the statement that $\delta_{\eta}\Seff=0$ for all these variations.

\section 3. Perturbation theory

From the point of view of perturbation theory,
the theories defined above are rather similar to lattice QCD
with Wilson fermions.
Important differences result from the use
of a relatively complicated lattice Dirac operator and from
the presence of the measure term, which gives rise
to additional gauge field vertices.
In this section we mainly address these issues,
while for the more common aspects of lattice per\-tur\-ba\-tion theory
the reader is referred to
refs.~[\ref{LesHouches}--\ref{RotheBook}], for example.

\subsection 3.1 Gauge fixing and BRS symmetry

When the gauge coupling $g_0$ is taken to zero,
the functional integral (2.12) is dominated by the 
field configurations
in the vicinity of the gauge orbit 
that passes through the trivial field $U(x,\mu)=1$.
The perturbation expansion essentially amounts 
to a saddle-point expansion about this orbit.
As usual the gauge degrees of freedom are first
eliminated by including a gauge-fixing term, 
and the gauge-fixed theory then has an exact BRS symmetry,
for any value of the lattice spacing
[\ref{LesHouches},\ref{BFM}].
Note that 
the effective action $\Seff$ does not interfere with this,
since it is gauge-invariant and of second order in the gauge coupling.

In the gauge-fixed theory
the gauge field is parametrized through
\equation{
  U(x,\mu)=\rme^{g_0aA_{\mu}(x)},
  \qquad
  A_{\mu}(x)=A^a_{\mu}(x)T^a.
  \enum
}
The integration variables are
then the components $A^a_{\mu}(x)$ of the gauge potential,
and the perturbation expansion is obtained straightforwardly 
by expanding all entries in the functional integral in powers of $g_0$.
Apart from the terms that derive from the effective action $\Seff$
or the fermion propagators in the Wick contracted product
$\{{\cal O}\}\subF$, the resulting Feynman rules are 
exactly as in the pure gauge theory 
and will not be discussed here.

\subsection 3.2 Expansion of the fermion propagator

The fermion propagator (2.13) may be written in the form
\equation{
  \{\psi(x)\psibar(y)\}\subF=
  S(x,y)P_{+}
  \enum
}
and our task is thus to expand the 
Green function $S(x,y)$ in powers of the gauge coupling. 
This has previously been described in 
refs.~[\ref{KikukawaYamada}--\ref{Capitani}], but
it may be worth while to briefly go through the 
main steps of this calculation to elucidate the 
structure of the free propagator and of the fermion-gauge-field vertices. 

In position space the Dirac operator $D$
is represented by a kernel $D(x,y)$ through
\equation{
  D\psi(x)=a^4\sum_y D(x,y)\psi(y).
  \enum
}
It suffices to work out the perturbation expansion
\equation{
  D(x,y)=\sum_{k=0}^{\infty}\kern1pt D^{(k)}(x,y),
  \enum
  \next{2ex}
  D^{(k)}(x,y)={g_0^k\over k!}\kern1pt
  a^{4k}\kern-3pt\sum_{z_1,\ldots,z_k}
  D^{(k)}(x,y,z_1,\ldots,z_k)^{a_1\ldots a_k}_{\mu_1\ldots\mu_k}
  A^{a_1}_{\mu_1}(z_1)\ldots A^{a_k}_{\mu_k}(z_k),
  \enum
}
of the Dirac operator, since the Green function is then obtained as usual
through the Neumann series
\equation{
  S(x,y)=S^{(0)}(x,y)-a^8\sum_{u,v}S^{(0)}(x,u)D^{(1)}(u,v)S^{(0)}(v,y)+
  \ldots
  \enum
}
Note that the kernels on the right-hand side of eq.~(3.5) are just the
bare vertices of the theory in position space with two fermion and
$k$ gauge field legs.

Starting from the definition (2.2) of the Dirac operator,
it is possible to compute these kernels analytically in momentum space.
Beyond the lowest orders the calculation
leads to increasingly complicated expressions, but
it is in principle straightforward and programmable.
An important simplification derives from the fact 
that the operator $A^{\dagger}A$ does not act on the Dirac indices 
at $g_0=0$. 
It is then not difficult to show that 
the free Dirac operator is given by
\equation{
  aD^{(0)}(x,y)=\int_{-\pi/a}^{\pi/a}
  {\rmd^4p\over(2\pi)^4}\,\rme^{ip(x-y)}
  \Bigl\{
  1-\bigl(1-\frac{1}{2}a^2\hat{p}^2-ia\dirac{\mu}\ring{p}_{\mu}\bigr)
  \lambda(p)^{-1/2}\Bigr\},
  \enum
  \next{3ex}
  \lambda(p)=
  1+\frac{1}{2}a^4\sum_{\mu<\nu}\hat{p}_{\mu}^2\hat{p}_{\nu}^2,
  \enum
}
where the standard notations
$\hat{p}_{\mu}=(2/a)\sin(ap_{\mu}/2)$ and 
$\ring{p}_{\mu}=(1/a)\sin(ap_{\mu})$ have been used.

To compute the fermion-gauge-field vertices, we expand the integrand 
in the integral representation
\equation{
  aD=1-\int_{-\infty}^{\infty}
  {\rmd t\over\pi}\, A\kern1pt(t^2+A^{\dagger}A)^{-1}
  \enum
}
in powers of the gauge coupling [\ref{KikukawaYamada}].
This yields a sum of products of
simple opera\-tors, and after passing to momentum space
the vertices are obtained in the form of in\-te\-grals of the type
\equation{
  \int_{-\infty}^{\infty}{\rmd t\over\pi}\,
  {P(q_1,\ldots,q_l)
  \over\left(t^2+\lambda(p_1)\right)\ldots\left(t^2+\lambda(p_r)\right)},
  \enum
}
where the numerator is a polynomial in the sines and cosines of
the incoming momenta and $p_1,\ldots,p_r$ are 
integer linear combinations of these. 
The integrals may finally be evaluated using the residue theorem.

It should be obvious from the above that the vertices
are analytic functions of the incoming momenta in 
a complex region around the Brillouin zone.
The kernel 
$D^{(k)}(x,y,z_1,\ldots,z_k)^{a_1\ldots a_k}_{\mu_1\ldots\mu_k}$
consequently falls off exponentially when the distance between
any two of its arguments becomes large.
Moreover the characteristic decay length 
is a fixed number in lattice units and hence
microscopically small from the point of view of the continuum limit.

The free lattice Dirac operator and the
fermion-gauge-field vertices are thus local,
as should be the case in a well-behaved theory.
In view of the general results of ref.~[\ref{Locality}],
this does not come as a surprise,
but having made it explicit what locality means in the present context 
will be helpful in the following.

\subsection 3.3 Expansion of the effective action

In the perturbation expansion of the functional integral (2.12),
the effective action $\Seff$ gives rise to additional 
(non-local) gauge field vertices
at the one-loop level. Up to an additive constant,
the expansion of $\Seff$ in powers of $g_0$ reads
\equation{
  \Seff=\sum_{k=2}^{\infty}\kern1pt
  {g_0^k\over k!}\kern1pt a^{4k}
  \kern-3pt\sum_{z_1,\ldots,z_k}
  V^{(k)}(z_1,\ldots,z_k)^{a_1\ldots a_k}_{\mu_1\ldots\mu_k}
  A^{a_1}_{\mu_1}(z_1)\ldots A^{a_k}_{\mu_k}(z_k).
  \enum
}
The vertices $V^{(k)}$ can be computed by differentiating 
eq.~(2.15) with respect to the gauge field,
but instead of $\delta_{\eta}$
another differential operator 
$\deltabar_{\eta}$ should better be used for this,
which acts on the gauge potential according to
\equation{
  \deltabar_{\eta}A_{\mu}(x)=\eta_{\mu}(x)
  \enum
}
(cf.~appendix A). 
The $k$-th order vertex is then obtained 
by applying this operator $k$ times to the effective action
and setting the gauge po\-ten\-tial to zero at the end of the 
calculation.

In terms of $\deltabar_{\eta}$, eq.~(2.15) assumes the form
\equation{
  \deltabar_{\eta}\Seff=
  -\Tr\{\deltabar_{\eta}D\hat{P}_{-}D^{-1}P_{+}\}+
  ig_0\L_{\bar{\eta}},
  \enum
  \next{2.5ex}
  \bar{\eta}_{\mu}(x)=
  \bigl\{1+\frac{1}{2}g_0a\Ad A_{\mu}(x)+\ldots\bigr\}\cdot\eta_{\mu}(x),
  \enum
}
where the higher-order terms in the curly bracket
are given explicitly in appendix~A. 
The differentiation of the trace term,
\equation{
  -\underbrace{\deltabar_{\eta}\ldots\deltabar_{\eta}}_
  {k-1\kern2pt{\rm times}}
  \left.\Tr\{\deltabar_{\eta}D\hat{P}_{-}D^{-1}P_{+}\}\right|_{A_{\mu}=0}=
  \noenum
  \next{1.5ex}
  \kern6em
  g_0^k\kern1pt a^{4k}
  \kern-3pt\sum_{z_1,\ldots,z_k}
  V_{\rm F}^{(k)}(z_1,\ldots,z_k)^{a_1\ldots a_k}_{\mu_1\ldots\mu_k}
  \eta^{a_1}_{\mu_1}(z_1)\ldots \eta^{a_k}_{\mu_k}(z_k),
  \enum
}
yields the fermion loop contribution to the vertices $V^{(k)}$.
As in the case of the fer\-mion propagator, the projector $\hat{P}_{-}$
on the left-hand side of this equation may be omitted.
The derivatives then act on the Dirac operator or 
its inverse, and as a result all
fermion one-loop diagrams with $k$ external gauge field lines
are generated.

In the next section the measure term will be obtained
in the form of a power series
\equation{
  \L_{\eta}=\sum_{k=4}^{\infty}\kern1pt
  {g_0^k\over k!}\kern1pt a^{4k+4}
  \kern-2pt\sum_{x,\ldots,z_k}
  L^{(k)}(x,z_1,\ldots,z_k)^{aa_1\ldots a_k}_{\mu\mu_1\ldots\mu_k}
  \noenum
  \next{1.0ex}
  \kern15em\times
  \eta^a_{\mu}(x)A^{a_1}_{\mu_1}(z_1)\ldots A^{a_k}_{\mu_k}(z_k),
  \enum
}  
with coefficients $L^{(k)}$ that are translation-invariant
and local (with exponentially decaying tails as in the case of 
the fermion-gauge-field vertices discussed above).
Since the series starts at $k=4$,
the vertices that derive from the measure term,
\equation{
  ig_0\underbrace{\deltabar_{\eta}\ldots\deltabar_{\eta}}_
  {k-1\kern2pt{\rm times}}
  \left.\L_{\bar{\eta}}\vphantom{\bigr\}}\right|_{A_{\mu}=0}=
  \noenum
  \next{1.5ex}
  \kern6em
  g_0^k\kern1pt a^{4k}
  \kern-3pt\sum_{z_1,\ldots,z_k}
  V_{\rm M}^{(k)}(z_1,\ldots,z_k)^{a_1\ldots a_k}_{\mu_1\ldots\mu_k}
  \eta^{a_1}_{\mu_1}(z_1)\ldots \eta^{a_k}_{\mu_k}(z_k),
  \enum
} 
only occur at the fifth and higher orders of the gauge coupling.
They are local linear combinations of the coefficients in eq.~(3.16),
properly symmetrized so as to comply with 
Bose symmetry.

\section 4. Determination of the measure term

We are now left with the task 
of determining the coefficients in eq.~(3.16) 
such that conditions (a) and (b) are satisfied
to all orders of the gauge coupling (cf.~sect.~2).
As explained in ref.~[\ref{NonAbelianChLGT}],
this is equivalent to solving a local cohomology problem in 
4+2 dimensions, but we shall not make use of this connection here
and instead construct the solution directly
through a recursive procedure,
assuming that the fermion representation of the gauge group
is anomaly-free.
The ``curvature"
\equation{
  \F_{\eta\zeta}=i\kern1pt
  \Tr\{\hat{P}_{-}[\delta_{\eta}\hat{P}_{-},\delta_{\zeta}\hat{P}_{-}]\},
  \enum
}
which appears on the right-hand side of the integrability condition
(2.17), plays an important r\^ole in this construction,
and its properties are thus worked out first.

\subsection 4.1 Gauge invariance, charge conjugation and the Bianchi identity

From the definition of the differential operator $\delta_{\eta}$
and the gauge-covariance of the projector to the left-handed fields,
it follows that $\F_{\eta\zeta}$ is invariant under gauge transformations
if $\eta_{\mu}(x)$ and $\zeta_{\mu}(x)$ 
are transformed like gauge-covariant local fields.

The lattice Dirac operator $D$ has the 
same charge conjugation pro\-per\-ties as the Dirac operator
in the continuum theory. 
In terms of the kernel $D(x,y)$, this means that its complex conjugate
is given by 
\equation{
 D(x,y)^{\ast}=BD(x,y)_{R\to R^{\ast}}B^{-1},
 \enum
}
where $B$ is a $4\times4$ matrix
such that $B\dirac{\mu}B^{-1}=\diracstar{\mu}{\ast}$.
The kernel of the projector to the left-handed fields 
transforms exactly in the same way, and since
the differential operator $\delta_{\eta}$ is real,
it follows that
\equation{
  \F_{\eta\zeta}=(\F_{\eta\zeta})^{\ast}=-(\F_{\eta\zeta})_{R\to R^{\ast}}.
  \enum
}
In particular, the curvature vanishes if the representation 
$R$ is real or pseudo-real.

Apart from being antisymmetric,
$\F_{\eta\zeta}$ also satisfies the Bianchi identity 
\equation{
  \delta_{\eta}\F_{\zeta\lambda}+
  \delta_{\zeta}\F_{\lambda\eta}+
  \delta_{\lambda}\F_{\eta\zeta}+
  a\F_{[\eta,\zeta]\lambda}+
  a\F_{[\zeta,\lambda]\eta}+
  a\F_{[\lambda,\eta]\zeta}
  =0.
  \enum
}
This can be proved in a few lines, using the commutator rule (A.6)
and $(\dirachat)^2=1$, which implies the vanishing of
$\Tr\{(\delta_{\eta}\dirachat)
(\delta_{\zeta}\dirachat)(\delta_{\lambda}\dirachat)\}$.

\subsection 4.2 Expansion of the curvature in powers of $g_0$

When the perturbation expansion of the Dirac operator 
is inserted on the right-hand side of eq.~(4.1),
a series of the form
\equation{
  \F_{\eta\zeta}=\sum_{k=0}^{\infty}\kern1pt
  {g_0^k\over k!}\kern1pt a^{4k+8}
  \kern-2pt\sum_{x,\ldots,z_k}
  F^{(k)}(x,y,z_1,\ldots,z_k)^{aba_1\ldots a_k}_{\mu\nu\mu_1\ldots\mu_k}
  \noenum
  \next{1ex}
  \kern15em\times
  \eta^a_{\mu}(x)\zeta^b_{\nu}(y)
  A^{a_1}_{\mu_1}(z_1)\ldots A^{a_k}_{\mu_k}(z_k)
  \enum
}
is obtained,
with local coefficients $F^{(k)}$ that are 
sums of products of the fermion-gauge-field
vertices. 
They are invariant 
under the adjoint action of $\group$
(i.e.~under con\-stant gauge transformations)
and change sign when the representation $R$ is replaced by $R^{\ast}$.

We now show that 
{\it all terms of order $k\leq2$ are equal to zero}\/
as a consequence of the anomaly cancellation condition
$\d{abc}=0$.
The proof is simple and starts with the observation that 
the link variables in the Dirac operator
only appear in the re\-pre\-sen\-tation $R$.
Taking the charge conjugation symmetry into account,
it follows from this and the general structure of the 
fermion-gauge-field vertices
that $F^{(k)}$ must be a linear combination of the 
tensors
\equation{
  \tr\{R(T^{c_1})\ldots R(T^{c_{k+2}})\}
  +(-1)^{k+1}\tr\{R(T^{c_{k+2}})\ldots R(T^{c_1})\},
  \enum
}
where $c_1,\ldots,c_{k+2}$ is any permutation of the indices
$a,b,a_1,\ldots,a_k$.
In particular, the leading order term $F^{(0)}$ is equal to zero, and 
the same is true
for $F^{(1)}$, because the tensor (4.6) is proportional to 
$\d{\kern1pt c_1c_2c_3}$ in this case.

For $k=2$ the vanishing of the tensor
can be proved by inverting the order of the generators 
in both traces simultaneously, using
\equation{
 [R(T^a),R(T^b)]=f^{abc}R(T^c).
 \enum
}
The commutator terms that are generated in this way 
do not contribute, since they
are proportional to the $\d{abc}$ symbol.
At the end of the calculation, the tensor is thus reproduced
with the opposite sign, which is only possible if it is equal to zero.

\subsection 4.3 Solution of the integrability condition to lowest order

On the left-hand side of the integrability condition (2.17), 
the differential operators
decrease the order of each term in the
series (3.16) since
\equation{
  \delta_{\eta}=g_0^{-1}\deltabar_{\eta}+\rmO(1)
  \enum
}
(cf.~appendix A). The first possibly non-zero term is thus of order
$g_0^3$ and if we define the lowest-order parts
of the measure term and the curvature through
\equation{
  \Lcheck_{\eta}=
  {1\over 4!}{\partial^4\over\partial g_0^4}
  \left.\vphantom{\biggl|}\L_{\eta}\kern2pt\right|_{g_0=0},
  \qquad
  \Fcheck_{\eta\zeta}=
  {1\over 3!}{\partial^3\over\partial g_0^3}
  \left.\vphantom{\biggl|}\F_{\eta\zeta}\kern2pt\right|_{g_0=0},
  \enum
}
the integrability condition at this order of the gauge coupling becomes
\equation{
  \deltabar_{\eta}\Lcheck_{\zeta}-\deltabar_{\zeta}\Lcheck_{\eta}
  =\Fcheck_{\eta\zeta}.
  \enum
}
Condition (a) must be satisfied too, and this
implies that $\Lcheck_{\eta}$ has to be invariant 
under {\it linearized gauge transformations}
\equation{
  A_{\mu}(x)\to A_{\mu}(x)+\partial_{\mu}\omega(x)
  \enum
}
and also under constant gauge transformations (where
both the gauge potential and $\eta_{\mu}(x)$ are rotated).

The following chain of arguments, which leads to a solution $\Lcheck_{\eta}$
of the problem, only makes use of the
locality, gauge invariance, homogeneity and antisymmetry
of $\Fcheck_{\eta\zeta}$ and of
the Bianchi identity 
\equation{
  \deltabar_{\eta}\Fcheck_{\zeta\lambda}+
  \deltabar_{\zeta}\Fcheck_{\lambda\eta}+
  \deltabar_{\lambda}\Fcheck_{\eta\zeta}=0
  \enum
}
that derives from eq.~(4.4).
For clarity, the construction
is broken up in four steps.

\vskip1ex
\noindent
{\bf 1.}
We first introduce a linear functional $\H_{\eta}$ through
\equation{
  \H_{\eta}=-\frac{1}{5}
  \left.\Fcheck_{\eta\lambda}\right|_{\lambda_{\mu}=A_{\mu}}=
  a^4\sum_x\eta^a_{\mu}(x)h^a_{\mu}(x),
  \enum
}
where the second equation defines the current $h_{\mu}(x)$.
Using the Bianchi identity and the homogeneity of 
$\Fcheck_{\eta\zeta}$, it is straightforward to show that
\equation{
  \deltabar_{\eta}\H_{\zeta}-\deltabar_{\zeta}\H_{\eta}
  =\frac{1}{5}
  \left\{2\kern1pt\Fcheck_{\eta\zeta}-\left(
  \deltabar_{\eta}\Fcheck_{\zeta\lambda}+\deltabar_{\zeta}
  \Fcheck_{\lambda\eta}\right)_{\lambda_{\mu}=A_{\mu}}\right\}=
  \Fcheck_{\eta\zeta},
  \enum
}
and $\H_{\eta}$ thus solves the leading-order form (4.10) of 
the integrability condition.

\vskip1ex
\noindent
{\bf 2.}
The current $h_{\mu}(x)$ itself may be gauge-dependent,
but its divergence
\equation{
  q(x)=\drvstar{\mu}h_{\mu}(x)
  \enum
}
can be proved to be invariant under linearized gauge transformations.
To this end let us consider two gauge variations 
$\eta_{\mu}(x)=\drv{\mu}\omega(x)$ and 
$\zeta_{\mu}(x)=\drv{\mu}\sigma(x)$.
Since the lowest-order part of the curvature is invariant under such 
variations, we have
\equation{
  \deltabar_{\lambda}\Fcheck_{\eta\zeta}=
  -\deltabar_{\eta}\Fcheck_{\zeta\lambda}
  -\deltabar_{\zeta}\Fcheck_{\lambda\eta}=0,
  \enum
}  
which proves that 
$\Fcheck_{\eta\zeta}$ is independent of the gauge potential
and hence equal to zero.
Recalling the definition (4.13), we now note that
\equation{
  \deltabar_{\zeta}\H_{\eta}=-\frac{1}{5}\kern1pt\Fcheck_{\eta\zeta}=0.
  \enum
} 
After substituting $\eta_{\mu}(x)=\drv{\mu}\omega(x)$ 
and performing a partial summation,
this is easily seen to be equivalent to the statement that
the divergence (4.15) is invariant under linearized gauge transformations.

\vskip1ex
\noindent
{\bf 3.}
From the above one concludes that $q(x)$ is 
a {\it topological field}, i.e.~it is local, invariant
under linearized gauge transformations and satisfies
\equation{
  a^4\sum_x\deltabar_{\lambda}q(x)=0
  \enum
}
for all variations $\lambda_{\mu}(x)$ of the gauge potential.
In lattice gauge theories with gauge group U(1),
it is known that 
any field with these properties can be written as a sum of a Chern polynomial
plus a topologically trivial term equal to the divergence of a
gauge-invariant local current [\ref{AbelianCohomology},\ref{FujiwaraEtAl}].

The theorem and its proof literally carry over to the present 
situation, where the components
$A^1_{\mu}(x),\ldots,A^n_{\mu}(x)$ behave
like independent abelian gauge fields.
We are actually dealing with a particularly simple case, 
because $q(x)$ is homogeneous 
in the gauge potential of degree $4$,
while the general Chern polynomial has degree $2$
in four dimensions.
The classification theorem thus implies
\equation{
  q(x)=\drvstar{\mu}k_{\mu}(x),
  \enum
}
where $k_{\mu}(x)$ is a local current that 
is invariant under linearized gauge transformations.
In refs.~[\ref{AbelianCohomology},\ref{FujiwaraEtAl}]
the current has been constructed algebraically
using a lattice version of the Poincar\'e lemma, and while
the resulting expression is rather complicated, 
it shows that $k_{\mu}(x)$ may be assumed to be homogeneous
of degree $4$ and to transform covariantly
under the adjoint action of the gauge group.

\vskip1ex
\noindent
{\bf 4.}
The lowest-order part of the measure term is now given by
\equation{
  \Lcheck_{\eta}=\H_{\eta}+\deltabar_{\eta}
  \Bigl\{\frac{1}{4}\kern0.5pt a^4\sum_x\,A^a_{\mu}(x)
  k^a_{\mu}(x)\Bigr\}.
  \enum
}
Since the second term has vanishing curvature, it is immediately clear
from eq.~(4.14) that the integrability condition in its leading-order
form (4.10) is satisfied. 
$\Lcheck_{\eta}$ is also local, homogeneous of degree 4 and 
invariant under constant gauge transformations.

To check the invariance of $\Lcheck_{\eta}$
under linearized gauge transformations,
we consider a gauge variation $\zeta_{\mu}(x)=\drv{\mu}\sigma(x)$
and note that
\equation{
  \deltabar_{\zeta}\Lcheck_{\eta}=
  -\frac{1}{5}\Fcheck_{\eta\zeta}
  +\deltabar_{\eta}
  \Bigl\{\frac{1}{4}\kern0.5pt a^4\sum_x\,\zeta^a_{\mu}(x)
  k^a_{\mu}(x)\Bigr\}.
  \enum
}
Use has been made here of the definition (4.13) and of
the gauge invariance of $\Fcheck_{\eta\lambda}$ and $k_{\mu}(x)$.
We already know that eq.~(4.10) holds, and the curvature 
$\Fcheck_{\eta\zeta}$ may thus be eliminated using this relation.
As a result one obtains
\equation{
  \deltabar_{\zeta}\Lcheck_{\eta}=
  \deltabar_{\eta}
  \Bigl\{-\frac{1}{4}\Lcheck_{\zeta}+\frac{5}{16}\kern0.5pt
  a^4\sum_x\,\zeta^a_{\mu}(x)
  k^a_{\mu}(x)\Bigr\}
  \noenum
  \next{2ex}
  \phantom{\deltabar_{\zeta}\Lcheck_{\eta}}=
  \deltabar_{\eta}
  \Bigl\{\frac{1}{4}\kern0.5pt
  a^4\sum_x\,\sigma^a(x)\left[\drvstar{\mu}h^a_{\mu}(x)-
  \drvstar{\mu}k^a_{\mu}(x)\right]\Bigr\}=0,
  \enum
} 
where the last equality follows from eqs.~(4.15) and (4.19).
$\Lcheck_{\eta}$ thus fulfils all conditions
to be an acceptable choice of the leading-order part of the 
measure term.

\subsection 4.4 Determination of the higher-order terms

The clue to the construction of the measure term at the
next-to-lowest order of the gauge coupling is the fact that
there exists a gauge-invariant local functional $\L^{(4)}_{\eta}$
whose lowest-order part coincides with $g_0^4\Lcheck_{\eta}$.
There are actually many such expres\-sions, and 
a particularly simple one is given in appendix C.
Once this is established, a subtracted measure term
and associated curvature may be defined through
\equation{
  \L'_{\eta}=\L_{\eta}-\L^{(4)}_{\eta},
  \enum
  \next{2ex}
  \F'_{\eta\zeta}=\F_{\eta\zeta}-
  \bigl\{\delta_{\eta}\L^{(4)}_{\zeta}-\delta_{\zeta}\L^{(4)}_{\eta}
  +a\L^{(4)}_{[\eta,\zeta]}\bigr\},
  \enum
}
in terms of which 
the integrability condition (2.17) assumes the form
\equation{
  \delta_{\eta}\L'_{\zeta}-\delta_{\zeta}\L'_{\eta}+
  a\L'_{[\eta,\zeta]}=
  \F'_{\eta\zeta}.
  \enum
}
The new curvature $\F'_{\eta\zeta}$ is of order $g_0^4$,
but has otherwise the same basic properties
(locality, gauge invariance, homogeneity, antisymmetry, Bianchi identity)
as $\F_{\eta\zeta}$. In particular, the lowest-order part
of $\L'_{\eta}$ can be determined by going through the steps
in the previous subsection again, with the obvious changes
that need to be made because 
the degree of homogeneity has increased by 1.

Evidently this procedure defines a recursion, which 
results in a series 
\equation{
  \L_{\eta}=\sum_{k=4}^{\infty}\,\L^{(k)}_{\eta},
  \enum
}
where $\L^{(k)}_{\eta}$
is of order $g_0^k$. The so constructed solution has 
all the required properties, and by expanding the terms in eq.~(4.26)
in powers of the gauge coupling, one finally obtains the coefficients
$L^{(k)}$.

\section 5. Further comments and results

\vskip-2.5ex

\subsection 5.1 Lattice symmetries

The imaginary part of the 
effective action should transform like a pseudoscalar
under lattice rotations and reflections,
but with the measure term $\L_{\eta}$ constructed in the 
previous section, this is not guaranteed.
We can, however, enforce the symmetry
by replacing $\L_{\eta}$
through the symmetrized expression 
\equation{
  {1\over2^44!}\sum_{\Lambda\in\rmO(4,\gz)}
  \det\Lambda
  \left.\L_{\eta}\right|_{U\to U^{\Lambda},\eta\to\eta^{\Lambda}}.
  \enum
}
Conditions (a) and (b) then are still fulfilled and 
the effective action 
has the desired transformation behaviour. 

The average (5.1) is taken over the group of integer orthogonal 
matrices $\Lambda$. They act on the 
lattice points and the gauge field in the usual way, while
in the case of the field $\eta_{\mu}(x)$
the transformation law is such that
\equation{
  \bigl[\rme^{ta\eta_{\mu}(x)}U(x,\mu)\bigr]^{\Lambda}=
  \rme^{ta\eta^{\Lambda}_{\mu}(x)}U^{\Lambda}(x,\mu).
  \enum
}
This implies a simple transformation behaviour of
the differential operator $\delta_{\eta}$ 
and the statement made above can then be 
proved straightforwardly.

\subsection 5.2 Uniqueness of the measure term

Conditions (a) and (b) do not fix the measure term uniquely,
but if we require that the lattice symmetries are preserved,
the difference $\Delta\L_{\eta}$ between any two 
solutions can be shown to be of the form 
\equation{
  \Delta\L_{\eta}=
  a^4\sum_x\delta_{\eta}\Omega(x), 
  \enum
}
where $\Omega(x)$ is a 
gauge-invariant, pseudoscalar local field.
Apart from the Chern monomials, which do not contribute
in perturbation theory due to their topological nature,
any field of this type has dimension greater than $4$.
Different choices of the measure term thus
amount to including further terms in the lattice action  
that are expected to be irrelevant in the continuum limit 
(up to finite renormalizations).

\subsection 5.3 Anomalous theories

If the fermion multiplet is anomalous, 
the expansion (4.5) of the curvature $\F_{\eta\zeta}$
starts at $k=1$ with a term proportional to $\d{abc}$
and the lowest-order part of the measure
term thus has to be a polynomial in the gauge potential of degree $2$.
The ar\-gu\-men\-tation in subsect.~4.3 then leads to 
a topological field $q(x)$ as before, but the field
now has degree 2 and can be topologically non-trivial.
From the results obtained in 
refs.~[\ref{NonAbelianChLGT},\ref{AbelianCohomology},\ref{FujiwaraEtAl}],
it is in fact possible to infer that 
\equation{
  q(x)=-{1\over192\pi^2}\d{abc}\epsilon_{\mu\nu\rho\sigma}
  T^aF^b_{\mu\nu}(x)F^c_{\rho\sigma}(x+a\hat{\mu}+a\hat{\nu})
  +\drvstar{\mu}k_{\mu}(x),
  \enum
}
where $F_{\mu\nu}(x)=\drv{\mu}A_{\nu}(x)-\drv{\nu}A_{\mu}(x)$
denotes the linearized gauge field tensor.
The construction of the measure term along the lines of sect.~4
thus breaks down at this point.

\subsection 5.4 Renormalizability

The lattice theories defined in this paper 
provide a gauge-invariant regularization of anomaly-free chiral 
gauge theories to all orders of the gauge coupling.
Moreover the construction preserves the
lattice symmetries (to the extent that this can be expected
in a chiral theory) and  
the propagators and basic vertices have all the 
required properties for 
the Reisz power counting theorem 
[\ref{ReiszPower},\ref{LesHouches}] to apply.

As a consequence there is little doubt that 
these theories are multiplicatively re\-normalizable, 
i.e.~it suffices to renormalize
the gauge coupling and the fields 
to be able to pass to the continuum limit.
Using techniques similar to those previously
employed in the case of lattice QCD
[\ref{ReiszRenormalization},\ref{ReiszRothe}],
it seems in  fact quite likely that a ri\-go\-rous proof of 
the multiplicative renormalizability 
can be given, even though the 
combinatorial aspects of the renormalization procedure
may have to be reconsidered, since 
the functional integral does not have the 
standard form with a local action and
field-independent integration measures for all fields.

\subsection 5.5 Right-handed fermions and Higgs fields

In theories with left- and right-handed fermions there 
are two multiplets of chiral fields, $\psi_L(x)$ and $\psi_R(x)$,
that transform according to some representations $R_L$ and $R_R$
of the gauge group.
Depending on which field the lattice Dirac operator $D$ acts,
the appropriate covariant difference operators should thus be used
in eq.~(2.3). The chirality constraints are then imposed 
as before, with the obvious changes.

Since the functional integrals over the left- and right-handed
fermions decouple, the total effective action is the sum
of the corresponding contributions. The same applies 
to the measure term $\L_{\eta}$, but 
it can be shown that the left- and right-handed parts
of the curvature $\F_{\eta\zeta}$ combine
to the purely left-handed expression (4.1) with the 
fermion representation $R$ given by
\equation{
  R=R_L\oplus(R_R)^{\ast}.
  \enum
}
In terms of this representation,
the anomaly cancellation condition is that $\d{abc}=0$,
and the construction of the measure term thus proceeds exactly
as in sect.~4.

It is now straightforward to add a Higgs field $\phi(x)$
that transforms according to the representation $R_L\otimes (R_R)^{\ast}$
of the gauge group. In particular, the obvious choice
\equation{
   \psibar_L(x)\phi(x)\psi_R(x)+\psibar_R(x)\phi(x)^{\dagger}\psi_L(x)
   \enum
}
for the Yukawa interaction is manifestly gauge-invariant and 
perfectly acceptable. 
An important point to note is that the introduction of the Higgs field
does not affect the measure term,
because the chiral projectors 
(and thus the fermion integration measure) do not refer to the Higgs sector.

\subsection 5.6 Calculation of electroweak processes

Lattice Feynman diagrams are relatively difficult
to evaluate, but having exact gauge invariance 
is a definite advantage when calculating electroweak amplitudes, which
may partly compensate for this.
If there are only few external lines,
it is quite clear that such computations are 
practically feasible at the one- and two-loop level.

\topinsert
\vbox{
\vskip0.1 true cm
 
\centerline{
\epsfxsize=1.8 true cm
\epsfbox{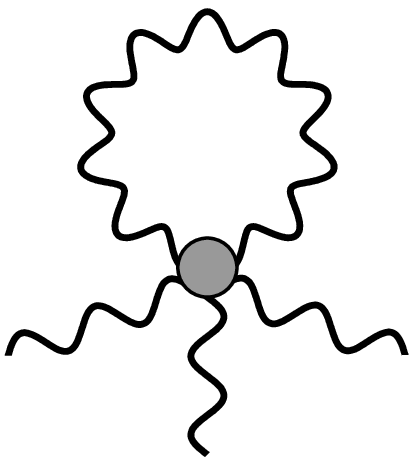}
}
\vskip0.0 true cm
\figurecaption{%
Feynman diagram involving the vertex $V_{\rm M}^{(5)}$ 
(shaded circle)
that derives from the measure term. The diagram appears at two-loop order,
while all other diagrams with such vertices
and less than four external lines are of higher order.
}
\vskip0.0 true cm
}
\endinsert

Although they could in principle be determined algebraically 
following the steps taken in sect.~4,
the vertices that derive from the measure term 
are not explicitly known at present.
For various reasons 
it seems rather unlikely, however, that they will ever be 
needed in the cases of interest.
In particular, 
these vertices only appear at the one-loop level 
and only at the fifth and higher orders of the gauge coupling.
Moreover they are proportional to a positive power of the lattice spacing
(a simple dimensional counting shows this)
so that at one-loop order they need not be included
if one is only interested in the continuum limit of the diagrams.

At the two-loop level the situation is more complicated, but 
dimension counting and symmetry considerations put strong constraints
on the amplitudes
to which the measure term can contribute.
The lowest-order (five-point) vertex, for example, 
is totally symmetric in the gauge group indices.
Together with the lattice symmetries, Bose symmetry and locality 
of the vertex, this suffices
to prove that the diagram shown in fig.~1 vanishes
in the continuum limit. One-particle irreducible 
diagrams at higher loop orders or with more than three external lines 
thus need to be considered to see 
a non-zero effect of the measure term.

\section 6. Concluding remarks

Regularizations of chiral gauge theories that preserve the gauge
symmetry must refer to the properties of the theory at the 
one-loop level,
because such a regularization can only exist if the gauge anomaly cancels.
For this reason simple schemes do not work out
and for many years the breaking of gauge invariance thus
appeared to be a necessary evil of any regularization of these theories.

In the lattice theories described in this paper 
the fermion integration measure has a non-trivial 
phase ambiguity that cannot be fixed consistently
if the fermion multiplet is anomalous.
The proper choice of the phase is an integral part of the definition 
of the lattice regularization, and the existence of the latter is thus
directly linked to 
the presence or absence of the anomaly.

Any anomaly-free chiral gauge theory can be regularized in this 
way, to all orders of the perturbation expansion,
but as is generally the case in lattice gauge theory,
the Feynman rules tend to be rather complicated.
Calculations of electro\-weak processes at the one- and two-loop level
may nevertheless be feasible, using algebraic manipulation programs,
the Reisz power counting theorem [\ref{ReiszPower},\ref{LesHouches}]
and a range of other tools 
to evaluate lattice Feynman diagrams.

An important question which has not been answered so far is whether
these lattice theories are multiplicatively renormalizable.
While there is little doubt that this is the case, 
a rigorous proof along the lines of 
refs.~[\ref{ReiszRenormalization},\ref{ReiszRothe}]
still needs to be given and would evidently be very welcome.

\vskip1ex
I am indebted to Tobias Hurth, Giuseppe Marchesini,
Raymond Stora and Hiroshi Suzuki for helpful discussions 
and to Peter Weisz for a critical reading
of a first version of this paper.

\appendix A

\vskip-2.5ex

\subsection A.1 Indices and Dirac matrices

Lorentz indices $\mu,\nu,\ldots$ are taken from the 
middle of the Greek alphabet and run from $0$ to $3$.
The symbol $\epsilon_{\mu\nu\rho\sigma}$ denotes the totally antisymmetric
tensor with~$\epsilon_{0123}=1$ and the conventions for the Dirac matrices are 
\equation{
  (\dirac{\mu})^{\dagger}=\dirac{\mu},
  \qquad
  \{\dirac{\mu},\dirac{\nu}\}=2\delta_{\mu\nu},
  \qquad
  \dirac{5}=\dirac{0}\dirac{1}\dirac{2}\dirac{3}.
  \enum
}
In particular, $\dirac{5}$ is hermitian and $(\dirac{5})^2=1$.

Fermion fields on the lattice carry a Dirac index
and a flavour index on which the gauge transformations act.
Indices $a,b,\ldots$ from the beginning of the Latin alphabet
are reserved for tensors that transform according to 
a tensor product of the adjoint representation 
of the gauge group.
Unless stated otherwise the Einstein summation convention is applied.

\subsection A.2 Gauge group

Without loss 
the gauge group $\group$ may be assumed to be 
a closed subgroup of U($N$) for 
some finite value of $N$ [\ref{Zelobenko}].
Its Lie algebra $\algebra$ is then a vector space
of anti-hermitian matrices and there exists 
a basis of generators $T^a$ ($a=1,\ldots,n$) such  that
\equation{
  \tr\{T^aT^b\}=-\frac{1}{2}\delta^{ab},
  \qquad
  [T^a,T^b]=f^{abc}T^c.
  \enum
}
With these conventions, the tensor $f^{abc}$
is real and totally antisymmetric.

The representation space of the adjoint representation of $\algebra$
is the Lie algebra itself, 
i.e.~the elements $X=X^aT^a$ of $\algebra$ are represented by linear 
transformations
\equation{
  \Ad X:\,\algebra\mapsto\algebra,
  \qquad
  \Ad X\cdot Y=[X,Y]\quad\hbox{for all}\quad Y\in\algebra,
  \enum
}
which is equivalent to 
\equation{
  (\Ad X\cdot Y)^a=f^{abc}X^bY^c
  \enum
}
in terms of the components of $X$ and $Y$.

\subsection A.3 Field variations

For any vector field $\eta_{\mu}(x)$ with 
values in $\algebra$ and compact support,
a first-order diffe\-ren\-tial operator $\delta_{\eta}$
acting on functions of the link variables may be defined
through
\equation{
  \delta_{\eta}f[U]={\rmd\over\rmd t}f[U_t]
  \kern-2.5pt\left.\vphantom{\Bigr\}}\right|_{t=0},
  \qquad
  U_t(x,\mu)=\rme^{ta\eta_{\mu}(x)}U(x,\mu).
  \enum
}
It is not difficult to show that $\delta_{\eta}f[U]$
is linear in $\eta_{\mu}(x)$ and that the identity
\equation{
  \delta_{\eta}\delta_{\zeta}-\delta_{\zeta}\delta_{\eta}
  +a\delta_{[\eta,\zeta]}=0
  \enum
}
holds if $\eta_{\mu}(x)$ and $\zeta_{\mu}(x)$ 
are independent of the gauge field.

Another kind of first-order differential operator $\deltabar_{\eta}$ 
acts on functions of the gauge potential according to
\equation{
  \deltabar_{\eta}g[A]={\rmd\over\rmd t}g[A+t\eta]
  \kern-2.5pt\left.\vphantom{\Bigr\}}\right|_{t=0}.
  \enum
}
These operators commute with each other and 
\equation{
  \deltabar_{\eta}f[U]=g_0\delta_{\bar{\eta}}f[U],
  \qquad
  U(x,\mu)=\rme^{g_0aA_{\mu}(x)},
  \enum
  \next{2.5ex}
  \bar{\eta}_{\mu}(x)=
  \biggl\{1+\sum_{k=1}^{\infty}{1\over(k+1)!}
  \kern1pt\bigl[g_0a\kern0.5pt\Ad A_{\mu}(x)\bigr]^k
  \biggr\}\cdot
  \eta_{\mu}(x),
  \enum
}
for any function $f[U]$ of the link variables.

\subsection A.4 Lattice derivatives

The forward and backward difference operators
$\drv{\mu}$ and $\drvstar{\mu}$ act on lattice functions
according to
\equation{
  \drv{\mu}f(x)={1\over a}\{f(x+a\hat{\mu})-f(x)\},
  \enum
  \next{1ex}
  \drvstar{\mu}f(x)={1\over a}\{f(x)-f(x-a\hat{\mu})\},
  \enum
}
where $\hat{\mu}$ denotes the unit vector in direction $\mu$.
They can be made gauge-covariant by including the appropriate
representation matrix of the link variables.
In the case of the fermion field the covariant forward
difference operator is given by
\equation{
  \nab{\mu}\psi(x)={1\over a}\bigl\{
  R[U(x,\mu)]\psi(x+a\hat{\mu})-\psi(x)\bigr\},
  \enum
}
and when $\omega(x)$ is a lattice field with values in $\algebra$, the
operator assumes the form
\equation{
  \nab{\mu}\omega(x)={1\over a}\bigl\{
  U(x,\mu)\omega(x+a\hat{\mu})U(x,\mu)^{-1}-\omega(x)\bigr\}.
  \enum
}
The covariant backward difference operator $\nabstar{\mu}$
is defined similarly.

\appendix B

In this appendix we prove that the 
effective action $\Seff$ is gauge-invariant if conditions (a) and (b)
are satisfied. 
Since we are only interested in the perturbative region,
it suffices to show that $\delta_{\eta}\Seff=0$ for all 
gauge variations (2.19) and all link fields in the vicinity of 
the trivial field $U(x,\mu)=1$.

We first note that
the gauge-covariance of the Dirac operator,
\equation{
  \delta_{\eta}D=[R(\omega),D],
  \enum
}
and eq.~(2.15) lead to the identity
\equation{
  \delta_{\eta}\Seff=\Tr\{R(\omega)(\hat{P}_{-}-P_{+})\}+i\L_{\eta}.
  \enum
}
The right-hand side of this equation is easily shown to vanish
at $U(x,\mu)=1$, using the explicit form (3.7) 
of the free Dirac operator in the first term
and the fact that $j_{\mu}(x)$ cannot depend on $x$
if the gauge field is translation-invariant.
To establish the gauge invariance of the effective action,
we then only need to prove that 
\equation{
  \delta_{\zeta}\L_{\eta}=i\kern1pt\Tr\{R(\omega)\delta_{\zeta}\hat{P}_{-}\}
  \enum
}
for arbitrary variations $\zeta_{\mu}(x)$ of the gauge field,
since this implies that the right-hand side of eq.~(B.2) is 
constant (and hence equal to zero).

Eq.~(B.3) is a simple consequence of the integrability condition
and the gauge-covariance of the current $j_{\mu}(x)$.
From the latter one infers that
\equation{
  \delta_{\eta}\L_{\zeta}+\L_{[\omega,\zeta]}=0,
  \enum
}
but the application of eq.~(2.17) is a bit tricky,
because $\eta_{\mu}(x)$ depends on the 
gauge field through the covariant difference operator in eq.~(2.19). 
We may, however, get around this problem by noting that
\equation{
  \delta_{\zeta}\L_{\eta}=
  \left\{\delta_{\zeta}\L_{\lambda}\right\}_{\lambda=\eta}
  +\L_{[\omega,\zeta]}+a\L_{[\zeta,\eta]}.
  \enum
}
Together with eq.~(B.4), the integrability condition then yields
\equation{
  \delta_{\zeta}\L_{\eta}=
  -i\kern1pt\Tr\{\hat{P}_{-}
  [\delta_{\eta}\hat{P}_{-},\delta_{\zeta}\hat{P}_{-}]\},
  \enum
}
which reduces to eq.~(B.3) when
the identities
\equation{
  \delta_{\eta}\hat{P}_{-}=[R(\omega),\hat{P}_{-}],
  \qquad
  \hat{P}_{-}\delta_{\zeta}\hat{P}_{-}\hat{P}_{-}=0,
  \enum
}
are inserted.

\appendix C

A local functional 
$\L^{(4)}_{\eta}$ with the required properties
can be obtained from the leading-order
part $\Lcheck_{\eta}$ of the measure term
by replacing the gauge potential in 
\equation{
  \Lcheck_{\eta}=
  {1\over 4!}\kern1pt a^{20}
  \kern-2pt\sum_{x,\ldots,z_4}
  L^{(4)}(x,z_1,\ldots,z_4)^{aa_1\ldots a_4}_{\mu\mu_1\ldots\mu_4}
  \eta^a_{\mu}(x)A^{a_1}_{\mu_1}(z_1)\ldots A^{a_4}_{\mu_4}(z_4)
  \enum
} 
through the expression
\equation{
  \Ahat^a_{\mu}(x,z)=
  {2\over a}\tr\bigl\{T^a
  \bigl[1-W(x,z)U(z,\mu)W(x,z+a\hat{\mu})^{-1}\bigr]\bigr\},
  \enum
}
where $W(x,z)$ denotes the ordered product of the link variables 
from $z$ to $x$ along the shortest path that first goes in 
direction $0$, then direction $1$, and so on.
From this definition and the invariance of
$L^{(4)}$ under the adjoint action of the gauge group,
it is obvious that $\L^{(4)}_{\eta}$ is gauge-invariant.
Moreover eq.~(C.2) implies
\equation{
  \Ahat_{\mu}(x,z)=
  g_0\bigl\{A_{\mu}(z)+\partial_{\mu}^z\omega(x,z)\bigr\}
  +\rmO(g_0^2)
  \enum
}
with $\omega(x,z)$ the ``oriented line sum" of the gauge potential 
from $z$ to $x$ along the path defined above.
Each term in the sum over $x$ in eq.~(C.1) is separately invariant
under linearized gauge transformations,
and $\L^{(4)}_{\eta}$ thus coincides with $g_0^4\Lcheck_{\eta}$
to leading order in the gauge coupling.


\ninepoint

\beginbibliography


\bibitem{BRS}
C. Becchi, A. Rouet, R. Stora, 
Commun. Math. Phys. 42 (1975) 127;
Ann. Phys. (NY) 98 (1976) 287

\bibitem{Becchi}
C. Becchi,
Lectures on the renormalization of gauge theories,
in: Relativity, groups and topology 
(Les Houches 1983), 
eds. B. S. DeWitt, R. Stora
(North-Holland, Amsterdam, 1984)

\bibitem{PiguetSorella}
O. Piguet, S. P. Sorella,
Algebraic renormalization: 
perturbative renormalization, symmetries and anomalies,
Lecture notes in physics m28
(Springer-Verlag, Berlin, 1995)


\bibitem{GrassiEtAl}
P. A. Grassi, T. Hurth, M. Steinhauser,
hep-ph/9907426

\bibitem{Jegerlehner}
F. Jegerlehner,
hep-th/0005255


\bibitem{Rome}
A. Borrelli, L. Maiani, G. C. Rossi, R. Sisto, M. Testa,
Phys. Lett. B221 (1989) 360;
Nucl. Phys. B333 (1990) 335

\bibitem{RomeReviewI}
M. Testa, 
The Rome approach to chirality,
in: Recent developments in non-perturbative quantum field theory
(Seoul 1997), eds. Y. M. Cho, M. Virasoro 
(World Scientific, Singapore, 1998)

\bibitem{BockEtAlI}
W. Bock, M. F. L. Golterman, Y. Shamir,
Nucl. Phys. (Proc. Suppl.) 63 (1998) 147 and 581;
Phys. Rev. Lett. 80 (1998) 3444;
Phys. Rev. D58 (1998) 034501

\bibitem{BockEtAlII}
W. Bock, K. C. Leung, M. F. L. Golterman, Y. Shamir,
Nucl. Phys. B (Proc. Suppl.) 83--84 (2000) 603;
hep-lat/9912025

\bibitem{RomeReviewII}
M. Testa, 
hep-lat/9912035


\bibitem{GinspargWilson}
P. H. Ginsparg, K. G. Wilson,
Phys. Rev. D25 (1982) 2649

\bibitem{PerfectDiracOperator}
P. Hasenfratz,
Nucl. Phys. B (Proc. Suppl.) 63A-C (1998) 53;
Nucl. Phys. B525 (1998) 401


\bibitem{IndexTheorem}
P. Hasenfratz, V. Laliena, F. Niedermayer,
Phys. Lett. B427 (1998) 125


\bibitem{NeubergerOperator}
H. Neuberger,
Phys. Lett. B417 (1998) 141;
{\it ibid.}\/ B427 (1998) 353


\bibitem{ExactSymmetry}
M. L\"uscher,
Phys. Lett. B428 (1998) 342


\bibitem{Locality}
P. Hern\'andez, K. Jansen, M. L\"uscher,
Nucl. Phys. B 552 (1999) 363


\bibitem{Narayanan}
R. Narayanan,
Phys. Rev. D58 (1998) 97501


\bibitem{Niedermayer}
F. Niedermayer,
Nucl. Phys. (Proc. Suppl.) 73 (1999) 105


\bibitem{AbelianChLGT}
M. L\"uscher,
Nucl. Phys. B549 (1999) 295

\bibitem{SuzukiAbelianChLGT}
H. Suzuki,
Prog. Theor. Phys. 101 (1999) 1147

\bibitem{NonAbelianChLGT}
M. L\"uscher,
Nucl. Phys. B568 (2000) 162

\bibitem{ReviewChLGT}
M. L\"uscher,
Nucl. Phys. B (Proc. Suppl.) 83--84 (2000) 34


\bibitem{AbelianCohomology}
M. L\"uscher,
Nucl. Phys. B 538 (1999) 515

\bibitem{FujiwaraEtAl}
T. Fujiwara, H. Suzuki, K. Wu,
Phys. Lett. B463 (1999) 63;
Nucl. Phys. B569 (2000) 643;
hep-lat/9910030


\bibitem{SuzukiAnomaly}
H. Suzuki,
hep-lat/0002009


\bibitem{KikukawaNakayama}
Y. Kikukawa, Y. Nakayama,
hep-lat/0005015


\bibitem{StoraI}
R. Stora,
Continuum gauge theories,
in: New developments in quantum field theory and statistical mechanics
(Carg\`ese 1976),
eds. M. L\'evy, P. Mitter (Plenum Press, New York, 1977)

\bibitem{StoraII}
R. Stora,
Algebraic structure and topological origin of anomalies,
in: Progress in gauge field theory 
(Carg\`ese 1983),
eds. G. 't Hooft et al. (Plenum Press, New York, 1984)

\bibitem{Zumino}
B. Zumino,
Chiral anomalies and differential geometry,
in: Relativity, groups and topology 
(Les Houches 1983), 
eds. B. S. DeWitt, R. Stora
(North-Holland, Amsterdam, 1984)

\bibitem{BaulieuI}
L. Baulieu,
Algebraic construction of gauge invariant theories,
in: Particles and Fields (Carg\`ese 1983),
eds. M. L\'evy et al. (Plenum, New York, 1985)

\bibitem{BaulieuII}
L. Baulieu,
Nucl. Phys. B241 (1984) 557;
Phys. Rep. 129 (1985) 1


\bibitem{BrandtEtAl}
F. Brandt, N. Dragon, M. Kreuzer,
Phys. Lett. B231 (1989) 263;
Nucl. Phys. B332 (1990) 224; 
{\it ibid.}\/ B332 (1990) 250

\bibitem{DuboisVioletteEtAl}
M. Dubois-Violette, M. Henneaux, M. Talon, C.-M. Viallet,
Phys. Lett. B267 (1991) 81;
{\it ibid.}\/ B289 (1992) 361


\bibitem{EriceLectures}
L. Alvarez-Gaum\'e,
An introduction to anomalies, 
in: Fundamental problems of gauge field theory (Erice 1985),
eds. G. Velo, A. S. Wightman (Plenum Press, New York, 1986)

\bibitem{Bertlmann}
R. A. Bertlmann, Anomalies in quantum field theory (Oxford University
Press, Oxford, 1996)


\bibitem{Witten}
E. Witten,
Phys. Lett. B117 (1982) 324;
Nucl. Phys. B223 (1983) 422

\bibitem{ElitzurNair}
S. Elitzur, V. P. Nair,
Nucl. Phys. B243 (1984) 205

\bibitem{BaerCampos}
O. B\"ar, I. Campos,
Nucl. Phys. B (Proc. Suppl.) 83--84 (2000) 594;
hep-lat/0001025


\bibitem{LesHouches}
M. L\"uscher,
Selected topics in lattice field theory,
in: Fields, strings and 
critical phenomena (Les Houches 1988), eds. E. Br\'ezin, J. Zinn-Justin 
(North-Holland, Amsterdam, 1989)

\bibitem{BFM}
M. L\"uscher, P. Weisz,
Nucl. Phys. B452 (1995) 213 and 234

\bibitem{MunsterMontvayBook}
I. Montvay, G. M\"unster,
Quantum Fields on a Lattice
(Cambridge University Press, Cambridge, 1994)

\bibitem{RotheBook}
H. J. Rothe,
Lattice gauge theories: an introduction, 2nd ed.
(World Scientific, Singapore, 1997)


\bibitem{KikukawaYamada}
Y. Kikukawa, A. Yamada,
Phys. Lett. B448 (1999) 265

\bibitem{AlexandrouEtAlI}
C. Alexandrou, H. Panagopoulos, E. Vicari,
Nucl. Phys. B571 (2000) 257

\bibitem{IshibashiEtAl}
M. Ishibashi, Y. Kikukawa, T. Noguchi, A. Yamada,
Nucl. Phys. B576 (2000) 501

\bibitem{AlexandrouEtAlII}
C. Alexandrou, E. Follana, H. Panagopoulos, E. Vicari,
hep-lat/0002010

\bibitem{Capitani}
S. Capitani, 
hep-lat/0005008


\bibitem{ReiszPower}
T. Reisz,
Commun. Math. Phys. 116 (1988) 81;
{\it ibid.}\/ 117 (1988) 79


\bibitem{ReiszRenormalization}
T. Reisz, 
Nucl. Phys. B318 (1989) 417


\bibitem{ReiszRothe}
T. Reisz, H. J. Rothe,
Nucl. Phys. B575 (2000) 255


\bibitem{Zelobenko}
D. P. \v{Z}elobenko,
Compact Lie groups and their representations
(American Mathema\-ti\-cal Society, Providence, 1973)

\endbibliography

\bye